\documentclass[10pt]{article}

\begin{document}

\title{ Everyone wants something better than $\Lambda$CDM}
  
\author
{Michael S. Turner \\
\normalsize Kavli Institute for Cosmological Physics\\
\normalsize University of Chicago, Chicago, IL  60637-1433\\
\\
\normalsize Department of Physics and Astronomy\\
\normalsize The University of California, Los Angeles\\
\normalsize Los Angeles, CA  90095-1547\\
\small 
email: mturner@uchicago.edu
\\
}
\maketitle

\vskip 0.2in

\begin{abstract}
The current cosmological paradigm, $\Lambda$CDM, is characterized its expansive description of the history of the Universe, its deep connections to particle physics and the large amounts of data that support it.  Nonetheless, $\Lambda$CDM's critics argue that it has been falsified or must be discarded for various reasons. Critics and boosters alike do agree on one thing:  it is the not the final cosmological theory and they are anxious to see it replaced by something better!  I review the status of $\Lambda$CDM, provide my views of the path forward, and discuss the role that the ``Hubble tension'' might play.

\end{abstract}
\vskip 0.1in

\noindent  $\Lambda$CDM is widely known for tracing cosmic history back to a very early inflationary beginning ($t \ll 10^{-6}$\,sec) through the formation of galaxies to today's accelerated expansion, as well as for its many successes, most notably agreement with precision measurements of the cosmic microwave background (CMB)  \cite{Turner2009}.  ($\Lambda$ is Einstein's cosmological constant, which is the placeholder for dark energy and accounts for today's accelerated expansion, and CDM stands for slowly-moving - ``cold'' - dark-matter particles, whose gravity holds all cosmic structures together.)

The three pillars of $\Lambda$CDM - dark matter, dark energy and inflation -  illustrate the deep connections between the very big, cosmology, and the very small, particle physics.  This paradigm shift took place at the end of the twentieth century \cite{Q2C} and remains a hallmark of both cosmology today \cite{EU} and particle physics \cite{EPP2024} today.  The trio is not only central to $\Lambda$CDM, but also they are among the big mysteries of both fields today.  

For all its successes, $\Lambda$CDM also has its critics who argue on a regular basis that it has been falsified or must be discarded for various reasons, from the existence of early, massive galaxies revealed by the James Webb Space Telescope to the theoretical shortcomings of inflation.  

So, is it a great achievement or failed theory?  As I will describe, critics and boosters alike agree on one thing, though for different reasons:  it needs to be replaced by something better.  I agree with that shared sentiment, and I would add that I hope that the successor to $\Lambda$CDM comes even closer to being a ``first principles'' theory than it is now.

\section*{Origins}

Around 1980, when I became a cosmologist, the hot big bang was the ``standard model'' of cosmology, having been given that moniker by Weinberg in his influential 1972 text, {\it Gravitation and Cosmology} \cite{HBB1972}, which attracted many physicists to the study of cosmology.  Before 1980, most cosmologists were astronomers and most learned their cosmology from Peebles's equally influential text, {\it Physical Cosmology} \cite{Peebles}.

The main successes of the standard model of cosmology included the existence of the cosmic microwave background, the big-bang synthesis of the lightest elements in the periodic table, and a general picture for describing how structure in the Universe - galaxies and the like - developed from small inhomogeneities in the matter density under the action of gravity.  

Back then, cosmology was data-starved:  direct evidence for the density inhomogeneities that seeded galaxies was lacking, the present expansion rate of the Universe  - the Hubble constant $H_0$ - was only known to within a factor of two, and only a small, nearby region of the Universe had been surveyed.

While the hot big-bang reigned for the next 20 years, the theoretical foundations for $\Lambda$CDM were developed:  the inflationary beginning, relic elementary particles as the dark matter, and the final piece, dark energy.  The ties between cosmology and particle physics became stronger and the ranks of cosmologists grew dramatically, with many newcomers from particle physics.

The evidence for the small density inhomogeneities (about a part in $10^5$) that seeded all structure was found by the Cosmic Microwave Background Explorer (COBE) satellite in 1992 \cite{COBEDMR}.  The discovery in 1998 that the expansion of the Universe was speeding up \cite{Hiz,SCP} and the CMB confirmation of inflation's prediction of a flat, critical density universe\footnote{Flat refers to a spatially uncurved universe; according to general relativity, a flat universe has a specific mean density, referred to as the critical density, $\rho_{\rm crit} = 3H_0^2/8\pi G \simeq 10^{-29}\,$g/cm$^3$, where $H_0$ is the present expansion rate and $G$ is the gravitation constant.} a few years later \cite{Langeetal2000} established $\Lambda$CDM as the consensus cosmology.  At the same time, cosmology went from a data-poor science to the flood of precision observations and measurements that characterizes it today \cite{Turner2022}.

\section*{From quantum fluctuations to accelerated expansion}

According to $\Lambda$CDM, during the very early inflationary phase the Universe grew exponentially in size, creating a big, smooth, flat region of spacetime large enough to contain all that we can see today and more \cite{Inflation1,Inflation2,Inflation3}.  The present description of inflation \cite{EU,Linde2007} involves a new scalar field - often called the inflaton - which, as far as is known, is unrelated to the one known scalar field, the Higgs field.

Quantum fluctuations in the scalar field responsible for inflation were blown up in physical size and became the tiny variations in the distribution of matter that seeded all the structures seen in the Universe today \cite{QF1,QF2,QF3,QF4}.   The seed perturbations in the matter density are seen in the anisotropy of the CMB first detected by COBE and studied since by other experiments.

The post-inflation Universe matches on to the hot big bang model with two critical additions to the first microsecond:  the creation of a tiny excess of matter over anti-matter, which when all the anti-matter and most of the matter annihilate leads to the small baryon-to-photon ratio (about $6\times 10^{-10}$) and the production of the slowly-moving dark-matter particles.   The details of the former, baryogenesis \cite{Baryo1,Baryo2}, and the identity of the dark-matter particle (or particles) both remain to be worked out \cite{DM1,DM2}.  

When the Universe was about 100,000 years old, the gravity of the dark matter acting upon the inflation-produced density perturbations began to assemble the rich array of structure seen in the Universe today, from galaxies to clusters of galaxies to walls and sheets of galaxies and regions that are devoid of galaxies (voids) \cite{CDM1985}.  

And finally, when the Universe was about 10 Gyr old, the repulsive gravity of dark energy, which accounts for 70\% of the Universe today, caused the expansion to begin accelerating.  

The defining feature of dark energy is its repulsive gravity,\footnote{In General Relativity, the ``gravity'' of stress energy that is elastic, i.e., has a negative pressure, can be repulsive rather than attractive.  The simplest example is the energy density associated with the vacuum, where the pressure is minus the energy density.}
 and quantum vacuum energy,\footnote{Quantum vacuum energy refers to the energy of the quantum vacuum.  It is the sum of the zero-point energies of all the quantum fields and it diverges, which is one of the big mysteries of particle physics.\cite{WeinbergRMP}.} mathematically the same as Einstein's $\Lambda$, is the placeholder for a deeper understanding of cosmic acceleration \cite{Stromlo,DECA}.
 
 $\Lambda$CDM not only provides a detailed account of the Universe from very close to its birth until today, it also makes a host of precise predictions that can be tested by astrophysical observations and laboratory experiments.  In the sense of Karl Popper, $\Lambda$CDM is a very strong theory because it makes many predictions can be used to falsify it.

\section*{Data rich, precision cosmology}

$\Lambda$CDM is a theory about the large-scale features of the Universe and the origin, evolution and distribution of mass and energy within it.  As such, its strongest support comes from the cosmic microwave background because the CMB probes the distribution of matter and energy at a simpler time, long before stars and galaxies.  

ESA's Planck \cite{Planck} and NASA's Wilkinson Microwave Anisotropy Probe \cite{WMAP} satellites and a host of ground-based CMB experiments \cite{Keck_BICEP,ACT,SPT} have measured the tiny (one part in $10^5$) variations of the CMB across the sky (anisotropy) and its small polarization (a few percent of the anisotropy!), on angular scales from 90 degrees to a fraction of an arcminute.  Each match the predictions of $\Lambda$CDM to percent-level precision.  

These measurements not only confirm the basic predictions of inflation - spatially-flat universe and almost scale-invariant density perturbations - but also provide the best determinations of the total matter density and that of ordinary matter (baryons) 30\% and 5\% respectively.  The almost 100-$\sigma$ difference between the total amount of matter and that in baryons is the air-tight evidence that dark matter is something other than atomic matter.  Of course, that is assuming that $\Lambda$CDM is correct.

Type Ia supernovae\footnote{These are the stellar explosions associated with the thermonuclear explosions of white dwarf stars that are pushed over the Chandrasekhar mass limit of 1.4 solar masses by accretion; they are excellent distance indicators because they are very bright and their luminosities can be standardized \cite{SNeReview}.}  in distant galaxies probe the expansion history, and measurements of thousands of such supernovae are consistent with the other 70\% of the Universe being dark energy in the form of quantum vacuum energy \cite{Pantheon,Union,DES}.

Using the light of stars to trace the mass distribution on galaxy scales and smaller is at best very difficult.  Baryons, which are but a small fraction of the total matter, do not trace the distribution of dark matter, and the stars that produce the light of galaxies account for only about 10\% of all the baryons and do not even trace the baryons faithfully.  

However, galaxies do map the distribution of matter shaped by the gravity of dark matter on large scales ($> 10\,$Mpc).  Galaxy surveys of the 3-dimensional distribution of tens of millions of galaxies out to distances of billions of light years have revealed a large-scale structure that is quantitatively consistent with that predicted by $\Lambda$CDM \cite{Planck,DES,DESI1,DESI2}.  

The wealth of precision cosmological data today leads to important crosschecks on $\Lambda$CDM.  The most impressive of which is the better than 1\% agreement of two very different determinations of the abundance of baryons.  The first is based upon nuclear physics when the Universe was seconds old and the production of deuterium during big-bang nucleosynthesis, and the second is based upon gravity-driven (acoustic) oscillations of baryons as they were falling into the dark matter potential wells when the Universe was 400,000 yrs old \cite{SchrammTurner,Cooke}.  This consistency also tests the underlying fundamental physics - General Relativity and the laws of nuclear physics - because if they were not constant in space and time, the two determinations would likely not agree.

\section*{The critics and their critiques }

So why is such successful theory under attack from so many directions?  Of course, that is the very nature of good science, and an expansive and successful theory attracts the most attention.  Further, stress-testing can uncover the loose ends that point the way to the next paradigm.

Astronomers interested in reconstructing the history of the Universe, from lumpy gas to stars and galaxies, are using the Webb and Hubble space telescopes as well as ground-based telescopes to reveal the birth of the first stars and galaxies.  $\Lambda$CDM is crucial to interpreting what they measure, e.g., converting redshift to age, angular size to physical size and flux to luminosity, which in turn allows them to better understand how the first galaxies and stars formed.

Some astronomers have tried to use their observations of the first galaxies to test $\Lambda$CDM.  This is challenging: $\Lambda$CDM is a theory about the distribution of mass and energy on the scales where gravity dominates; connecting these predictions to the light produced by stars involves the interplay of gravity, electromagnetic and nuclear interactions \cite{Primack}, what many refer to as messy ``gastro-physics.''

Nonetheless, a spate of papers based upon the early results from the Webb telescope claimed to overturn $\Lambda$CDM because of the overabundance of massive galaxies at redshifts greater than ten \cite{KillLCDM1,KillLCDM2}.   Add to the concerns mentioned above, the early star formation revealed by the Webb telescope is wildly different from what we see today.  By now, the misplaced claims of troubles for $\Lambda$CDM have all but disappeared.

It would be wonderful if $\Lambda$CDM, or its successor, could make very specific predictions about how the Universe lit up so that observations of the distribution of stars and galaxies on galactic scales could be used to test $\Lambda$CDM.  However, that is a very tall order, and in my opinion, about as unlikely as biochemists creating a first principles model of consciousness.  

\section*{Cracks in the foundations?}

A different set of objections involve the theoretical underpinnings of $\Lambda$CDM.  Cold dark matter, $\Lambda$, and scalar-field inflation are place holders for more fundamental explanations of dark matter, cosmic acceleration, and the origin of the large-scale smoothness of the Universe and the density perturbations that seeded all cosmic structure.  Very successful placeholders but not satisfying to those looking for a fundamental, or ``Lagrangian-level,'' explanation for each.  

By identifying the dark-matter particle and its place in a grander particle physics theory, by providing a fundamental explanation for cosmic acceleration, and by tying inflation to a specific scalar field, $\Lambda$CDM could be ``upgraded'' to a fundamental theory akin to the Standard Model of particle physics.  

Others see larger problems that point to even more radical revisions that may be needed \cite{Penrose}.  Steinhardt \cite{InflationDebate} and his collaborators argue that inflation is so fundamentally flawed that it is not even a theory and that it needs to be replaced, e.g., by their cyclic model \cite{Cyclic}.  They point to inflation's apparent prediction of a multiverse structure for the Universe \cite{Vilenkin}, which, according to them, makes specific predictions about things that can measured - even in one piece of the multiverse - impossible.  

Here, I say apparent prediction of a multiverse, because the current version of inflation is not a complete theory by any stretch of the imagination.  A fundamental version of inflation might address the multiverse issue or even make it moot by not predicting a multiverse at all.  I would agree that inflation is at present more of a paradigm than a theory, and whether or not it needs to be discarded or merely fleshed out is an open question.

Other, even more technical, issues have been raised.  For example, the tremendous expansion of the Universe during inflation stretches quantum effects on extremely tiny scales, where quantum gravitational effects should be important and are not well understood, to large enough size that should be able to influence the Universe and have observable effects.  Could their effects radically alter the predictions of $\Lambda$CDM \cite{subplanckian}?  

The most intriguing concern is the so-called swampland conjecture \cite{swampland}:  namely, that string theory is inconsistent with inflation as formulated and with the current epoch of cosmic acceleration if it is due to $\Lambda$.  Of course, whether or not string theory is correct is still to be determined!

The expectations for a cosmological theory have risen as particle physicists have joined the ranks of cosmologists.  Before the hot big bang became the standard model of cosmology, Sandage called cosmology, ``a search for two numbers'' (the expansion rate and the deceleration parameter), a purely phenomenological description of the Universe \cite{Sandage}.  

The hot big bang had its roots in General Relativity, nuclear and atomic physics, and the foundation of $\Lambda$CDM is General Relativity and the Standard Model.  Today, many - including myself - aspire for a fundamental theory of the Universe on a par with the Standard Model of particle physics.  

\section*{MOND and dark matter}

The very idea of dark matter is a flash point for some astronomers.  In 1983, shortly after the flat rotation curves of galaxies measured by Rubin \cite{Rubin} and others made a strong case for the dark-matter halos of galaxies \cite{FaberGallagher}, Milgrom \cite{Milgrom} pointed out that flat rotation curves could be explained by modifying the familiar Newtonian dynamics, $F=ma$.  His idea is known as MOND, for MOdified Newtonian Dynamics.  

Milgrom showed that the need for dark matter in galaxies occurs at a fixed acceleration, around $10^{-8}$\,cm/sec$^2$.  And if at accelerations below this, the force needed to keep stars orbiting the centers or their galaxies were proportional to acceleration squared, not acceleration, there would be no need for dark-matter halos in galaxies.  

The fact the need for dark matter in galaxies occurs at a universal acceleration is remarkable and can be accounted for in $\Lambda$CDM \cite{MilgromLaw}.  Further, the adherents of MOND have not been successful in turning MOND into a relativistic theory of gravity that can match the many successes of $\Lambda$CDM beyond galactic rotation curves, let alone  make new predictions that can decisively test it.  Moreover, the evidence for dark matter today goes well beyond galactic rotation curves, and now includes the dark matter in clusters of galaxies, the CMB determination of the matter and baryon densities, and the necessity of dark matter for explaining the large-scale structure of the Universe.

\section*{Cold, warm, fuzzy or self-interacting}
There is a simpler question about dark matter:  is it cold and without dynamically important interactions beyond that of gravity?  If so, the central dark-matter density in galaxies should be ``cuspy'' and the number of small dark matter clumps should rise with decreasing mass \cite{CDMproblems}.  There is evidence that galactic rotation curves are not as cuspy as CDM predicts, and to date, not enough small clumps have been found.  

To address these potential problems, modifications to the properties of the dark matter particle have been suggested:  that it is warm \cite{WDM} or fuzzy \cite{fuzzyDM} rather than cold or has dynamically important interactions with other dark-matter particles (self-interacting) \cite{selfinteracting}.

However, the situation is murky.  The inner-most portions of the rotation curves of faint dwarf galaxies, where dark matter alone can be probed, are the hardest to measure; and, the baryons that are concentrated there and can have significant effect on the distribution of dark matter.  The smallest dark-matter clumps are expected to be very faint and difficult to see. Moreover, recent galaxy surveys have found rising numbers of faint, low-mass dwarf galaxies in the Local Group \cite{DESdwarfs1,DESdwarfs2} and gravitational lensing has revealed clumps in galactic halos \cite{DMLumps}.  

While, these galactic-scale problems may be indicative of serious problems with CDM and the need for radical change, the case for such is not yet compelling \cite{CDMproblems}.

Discovering the dark-matter particle (or particles!) would put most of these concerns to rest.  But, despite great effort, that has not happened.  The focus has been on a particle of mass 100 to 1000 times that of the proton, whose interactions with ordinary matter are of roughly the strength of the usual weak interactions, often referred to as the WIMP, for Weakly Interacting Massive Particle.

These two features make WIMPs detectable in three different ways:  creation in particle collisions at the Large Hadron Collider at CERN, direct detection of Milky Way halo dark-matter particles in very sensitive, multi-ton underground detectors, and through the annihilation of halo dark matter into positrons, antiprotons or gamma rays.  All three methods have come up empty, and theorists are busy exploring new possibilities for the dark-matter particle(s) \cite{DM1,DM2}.

Non-baryonic dark matter is central to $\Lambda$CDM, as well as to both astrophysics and particle physics.   We already have evidence that a small part of the mass density is not baryons - the neutrinos left over from the big bang with their tiny masses account for between 0.1\% and 0.15\% of energy content of the Universe \cite{DESI2}, compared to 0.5\% for stars, 5\% for baryons and 30\% total for matter.  The story may well be more complicated than just CDM.

\section*{No cosmic acceleration?}

The last group of dissenters focus on cosmic acceleration. Within the usual framework of a nearly isotropic and homogeneous Universe, the evidence for cosmic speed up is iron-clad.  

However, if you drop the assumption of homogeneity and posit that we live precisely in the middle of a large isotropic but inhomogeneous region with just the right density profile, the interpretation of the distance-redshift measurements could be made consistent with a slowing expansion \cite{Enqvist}.  

Others have argued the inhomogeneity that exists on small scales - galaxies, clusters of galaxies, and so on - could give rise to an apparent cosmic acceleration, when such inhomogeneity is properly considered within General Relativity.  But they have been unable to describe how such a large effect would arise \cite{Buchert}.

Last year, the plot thickened.  The Dark Energy Spectroscopic Instrument (DESI) made percent-level measurements of distances out to redshift four that probe the expansion history of the Universe.  By combining their results with CMB and supernovae measurements, the DESI collaboration found evidence at greater than 3-$\sigma$ significance that dark energy is evolving, which is inconsistent with quantum vacuum energy \cite{DESI1,DESI2}.   If this result holds up, it could shed light on the reality of cosmic acceleration and/or the mechanism causing it.

\section*{What does better look like?}

$\Lambda$CDM is a remarkable theory.  It describes in detail the evolution of the Universe from quantum fluctuations and inflation to a Universe filled with galaxies moving away from one another at an accelerating rate, and, it makes a host of predictions, which have been verified by a wealth of precision data and continue to test it. 

While its critics have raised concerns, none of them at present rises to the level of falsifying it.  Some of the issues do however point to the loose ends and even bigger questions that the next cosmological paradigm should aspire to address, including the origin of the Universe itself.  

I believe that the common thread in all of this is the hope of replacing $\Lambda$CDM with a theory that is more ``first principles'' and less phenomenological. 

$\Lambda$CDM has a first principles foundation, General Relativity and the Standard Model of particle physics, and phenomenological add ons, dark matter, dark energy and inflation.  A fully first principles successor {\it could} be the successor to the Standard Model, which fully incorporates dark matter, dark energy and inflation plus General Relativity, or thinking much more boldly, it {\it could} be the quantum theory that marries the Standard Model and General Relativity and explains the origin of space, time and the Universe.  The stakes are very high and we could be on the verge of a major advance in our understanding the Universe and the laws that govern it.

I would dare say that $\Lambda$CDM is much more than any cosmologist - certainly myself - hoped for in 1980.  But now, it is much less than cosmologists are willing to settle for \cite{Turner2018}.  So how do we get to this grander theory that we all want?

\section*{The Hubble tension}

The so-called Hubble tension \cite{SHOES,CCHP,WLF,MarcAdam} could be the path forward.  The tension is in essence an end-to-end test of $\Lambda$CDM:  precision measurements of the expansion rate of the Universe today find $H_0 =73 \pm 1$\,km/s/Mpc, while even more precise measurements involving the cosmic microwave background arrive at $H_0 = 66.6 \pm 0.5$\,km/s/Mpc.   

Both measurements could be correct if $\Lambda$CDM is incomplete or wrong.  That is because the CMB measurement of $H_0$ is actually a measurement of the expansion rate when the Universe was 380,000 yrs old, extrapolated forward 14 billion years by $\Lambda$CDM,\footnote{The actual expansion rate determined by CMB measurements, before it is extrapolated to the present, is $1.6 \times 10^6$\,km/sec/Mpc.} and it is being compared with a direct measurement of the expansion rate today.  $\Lambda$CDM brings the two measurements together.

If $\Lambda$CDM is missing something, it could reveal itself in the Hubble constant  discrepancy.  

On the other hand, percent-level precision is no mean feat in cosmology, and the resolution could involve subtle systematic errors in one or both determinations of $H_0$.  Already, these two measurements provide an end-to-end test of $\Lambda$CDM, which it passes at the 10\% level, and if systematic errors are found that bring the measurements into agreement, the test would be even more impressive.

Cosmology faced a similar situation in the mid-1990s.  Before $\Lambda$, inflation + cold dark matter had many successes, but there were also problems: not enough dark matter to account for the flat Universe predicted by inflation, an age/Hubble-constant conflict, and disagreements with measurements of the large-scale structure of the Universe.  Inflation + CDM was on the brink of losing out to a competing theory, cosmic strings \cite{CS}.

In 1995, several of us pointed out that adding $\Lambda$ would solve all the problems \cite{KraussTurner,PJSJPO,MarcyHistory}.  A few years later, the discovery of cosmic acceleration made $\Lambda$ a reality and $\Lambda$CDM soon became the consensus cosmology.

It would be exciting if history repeated itself.  However, to me, the modifications to $\Lambda$CDM suggested to resolve the Hubble tension seem ad hoc and are not yet compelling.  

Will the path forward be evolution - discovering the dark matter particle, understanding dark energy at a more fundamental level, and finding a standard model of inflation - or revolution - for example, replacing dark matter and dark energy with a new theory of gravity?  The triggering event could be a surprising observation, a discrepancy that won't go away, or a theoretical breakthrough.  When this happens, remains to be seen.

Linde, one of the originators of inflation, more than once said, you can only kill inflation with a better theory.  While I found his statement jarring the first time I heard it, I think it has some relevance here.  

Given $\Lambda$CDM's many successes and the absence of a competitor that can match even a few of them, it is too early to discard it.  And, because the body of data that supports it is so extensive, its successor will first have to make itself look like $\Lambda$CDM before it reveals its exciting new features.

\vskip 0.2in
\noindent{I thank the organizers and participants of the 2025 Lake Como School on Dark Matter, Dark Energy, and the Cosmological Tensions for many lively conversations about the virtues and deficiencies of $\Lambda$CDM. The many stimulating conversations with Tommaso Treu, Richard Ellis and Martin Rees also helped to inform my thinking.  I thank others, including Alexander Fetter and Wendy Freedman, for insightful comments on an early draft.}


\bibliographystyle{unsrt}
\bibliography{pnas_LCDM}

\end{document}